\documentclass[
	a4paper, 
	10pt, 
	unnumberedsections, 
	twoside, 
]{LTJournalArticle}

\usepackage{tikz}
\newcommand*\circled[1]{\tikz[baseline=(char.base)]{
            \node[shape=circle,draw,inner sep=0.7pt] (char) {#1};}}

\runninghead{} 

\footertext{\textit{RISC-V Summit Europe, Munich, 24-28th June 2024}} 

\title{Design, Implementation and Evaluation of the SVNAPOT Extension on a RISC-V Processor} 

\author{%
	Nikolaos-Charalampos Papadopoulos\textsuperscript{1}\thanks{Corresponding author: \href{mailto:ncpapad@cslab.ece.ntua.gr}{\tt ncpapad@cslab.ece.ntua.gr} \\ This work was partially supported by the EU funded Vitamin-V project \#101093062.}, Stratos Psomadakis\textsuperscript{1}, Vasileios Karakostas\textsuperscript{2}, \\ Nectarios Koziris\textsuperscript{1} and Dionisios N. Pnevmatikatos\textsuperscript{1}
}

\date{\footnotesize\textsuperscript{\textbf{1}}National Technical University of Athens\\ \textsuperscript{\textbf{2}}National and Kapodistrian University of Athens}

\renewcommand{\maketitlehookd}{%
    \vspace{-0.5cm}
	\begin{abstract}
    \vspace{-0.2cm}
		\noindent The RISC-V SVNAPOT Extension aims to remedy the performance overhead of the Memory Management Unit (MMU), under heavy memory loads. The  Privileged Specification defines additional Natural-Power-of-Two (NAPOT) multiples of the 4KB base page size, with 64KB as the default candidate. In this paper we extend the MMU of the Rocket Chip Generator, in order to manage the collocation of 64KB pages along with 4KB pages in the L2 TLB. We present the design challenges we had to overcome and the trade-offs of our design choices. We conduct a preliminary sensitivity analysis of the L2 TLB with different configurations/page sizes. Finally, we summarize on techniques which could further improve memory management performance on RISC-V systems.

	\end{abstract}
}

\begin{document}

\maketitle 


\section{Design}
\begin{figure}[ht] 
\centering
\includegraphics[scale=0.9]{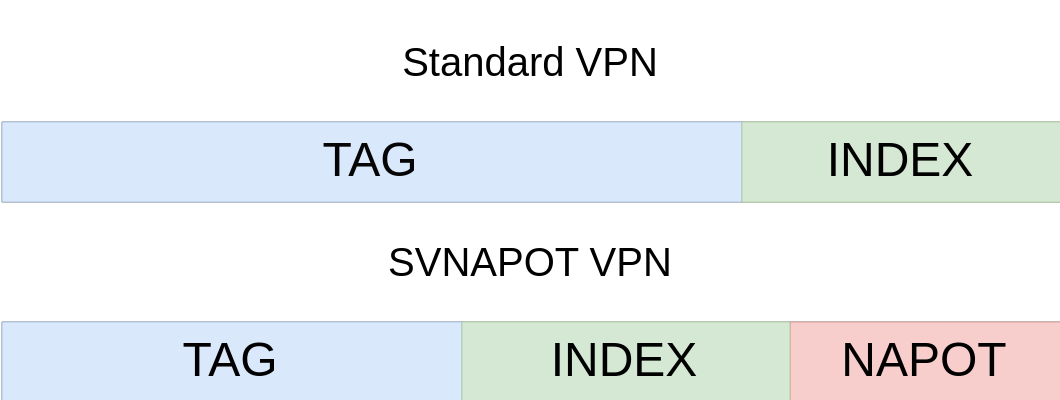} 
\vspace{-0.1cm}
\caption{VPN Partitioning}
\label{figure_svnapot}
\end{figure}

\vspace{-0.3cm}
We extend the L2 TLB of Rocket\cite{asanovic01} and add support for SVNAPOT\cite{waterman02}, targeting the Page-based 39-bit Virtual Memory systems (\textit{sv39}) scheme. We choose to collocate both 4KB and 64KB in the same structure, in order to be able to take advantage of every L2 TLB entry. We modify each entry, and add an N bit which if set, it indicates that the TLB entry holds a 64K page. Structures that collocate different page sizes exhibit an indexing problem: For each TLB lookup we do not know the page size and thus cannot determine the set (index) it may reside in. The common methods of resolving this problem are (1) multiple lookups for different page sizes, (2) Virtual-Page-Number (VPN) partitioning and (3) companion TLBs. We choose VPN partitioning in order to avoid wasting cycles in multiple lookups. We ignore the last 4 bits of the VPN, and select the next log2(\#sets) bits (figure \ref{figure_svnapot}) as INDEX.  We modify the following operations in order to be able to handle two page sizes:

\subsubsection{Insert} Set the N bit according to the PTE returned by the Page Table Walker (PTW).

\subsubsection{Lookup} If N=1 we found a 64KB page $\rightarrow$ return (PPN $\gg$ 4) + NAPOT\_OFFSET. If N=0 return PPN.

\subsubsection{Flush} Locate INDEX and flush the whole set.




\vspace{-0.3cm}
\vspace{-1cm}










\section{Methodology}
\vspace{-0.3cm}
\subsubsection{Software}
We use an unmodified v6.8-rc5 Linux kernel, which implements the SVNAPOT extension and supports allocations for 4KB and 64KB pages. Since 64KB SVNAPOT pages are not yet transparently supported, we opt to pre-allocate 8GB of 64KB pages. For 64KB allocations, the kernel includes the MAP\_HUGE\_64KB flag for the mmap() system call. 

\subsubsection{Hardware}
The system under test has a single Rocket core with an unmodified 32-entry DTLB and an 8-entry PTW Cache. We set up 4 different evaluation scenarios (Table \ref{tableConf}) along 2 different L2 TLB configurations, a (1) 4-way set-associative and (2) 16-way set-associative organization. Both configurations have 1024 entries.

\subsubsection{Simulation}
We employ FireSim\cite{firesim}, a fast FPGA-accelerated hardware simulator with a Xilinx Alveo U280 board that supports 16GB of DRAM.

\vspace{-0.3cm}

\begin{table}[ht]
\centering
\begin{tabular}{ c c c } 
\midrule
\textbf{Conf. No} & \textbf{Associativity} & \textbf{Page Size}  \\ 
\midrule
\circled{1} & 4-way & 4KB \\ 
\circled{2} & 16-way & 4KB \\
\circled{3} & 4-way & 64KB  \\
\circled{4} & 16-way & 64KB \\

\midrule
\end{tabular}
\caption{Evaluated Configurations}
\label{tableConf}
\end{table}
\vspace{-0.7cm}

\vspace{-0.3cm}
\section{Evaluation}
\vspace{-0.3cm}
In order to evaluate our design, we compile a comprehensive sensitivity analysis using a microbenchmark that targets different memory requirements and access patterns. This TLB-stress microbenchmark spans a range of memory chunks from 4KB to 256MB, with a linear access pattern \circled{A} and a random access pattern \circled{B}. For each iteration, TLB-stress warms up the TLBs and then performs 1M memory accesses with 4KB step in order to stress the system. We also properly modify and test the hashjoin benchmark, in order to be able to use 64KB page allocation.

\begin{figure}[ht] 
\vspace{-0.4cm}
\centering
\includegraphics[scale=0.21]{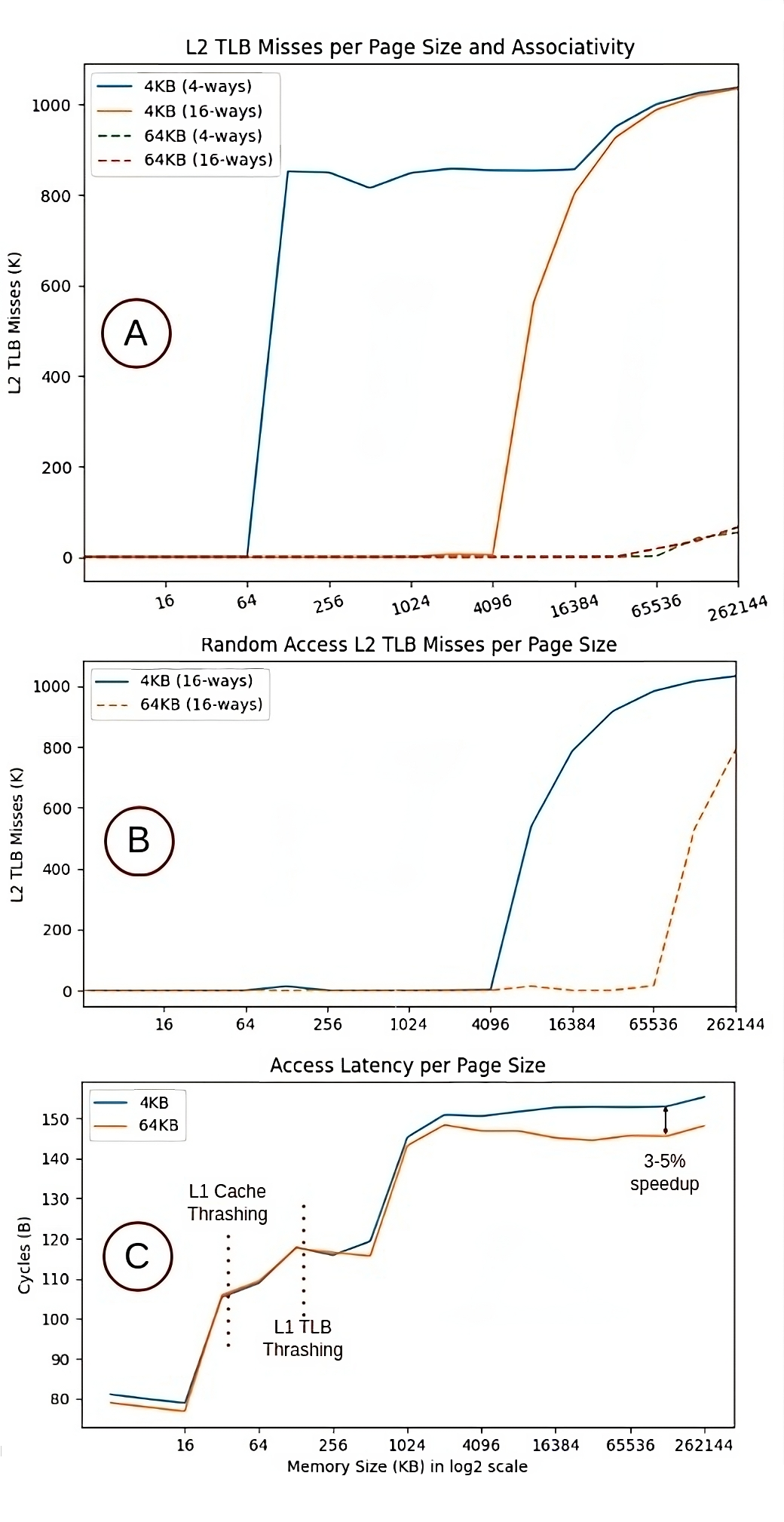} 
\vspace{-0.7cm}
\caption{Experimental Results. The x axis in all diagrams reports Memory Size (KB) in log2 scale.}
\label{figure_combo}
\vspace{-0.7cm}
\end{figure}

In Figure \ref{figure_combo}, we exhibit the results we gathered with TLB-stress. In \circled{A}, TLB-stress performs a linear page stride for all configurations. We observe that in \circled{1} the L2 TLB gets thrashed after the 64KB mark. This happens due to the 4-bit shift in the TLB index that causes conflicts in every set; ignoring the last 4-bits means the index changes in a 16-page interval. Also, until the 128KB mark most lookups are serviced by the L1 DTLB. Early L2 thrashing is remedied in \circled{2}, in which the L2 TLB exhibits the expected 4MB reach. Configurations \circled{3}, \circled{4} showcase a rapid decrease in misses (misses reached up to 70K). In \circled{B}, TLB-stress performs random accesses for \circled{2}, \circled{4}. We observe that both configurations present zero misses until the 4MB and 64MB mark for \circled{2} and \circled{4} respectively, with \circled{4} indicating 16$\times$ larger reach. Finally, in \circled{C} we demonstrate latency results in terms of cycles  during a page stride for configurations \circled{2} and \circled{4}. Following the 16KB mark, \circled{4} outperforms \circled{2} by 3-5\%.

Regarding hashjoin, we tested \circled{2} and \circled{4}. Using 64KB pages, hashjoin exhibited 4.2\% speedup in terms of cycles while reducing L2 TLB misses by 14.2\%.

\vspace{-0.2cm}
\subsubsection{Latency}
The cost of one 64KB page PTW (TLB refill) is equivalent to (and saves up to) 16$\times$4KB PTWs. L2 TLB Lookup latency is not affected and is identical for both page sizes (3 cycles to report hit/miss).

\vspace{-0.2cm}
\subsubsection{Area}
Negligible, 1.1\% overhead in the L2 TLB structure (1024 N bits), plus simple logic in order to distinguish between 4KB and 64K pages. We do not add any other persistent elements such as registers/SRAMs.

\vspace{-0.2cm}
\subsubsection{Power}
A PTW is a costly operation since multiple memory reads are needed in order to locate the PPN. This can get worse for schemes that support extra page table levels, such as \textit{sv48} and \textit{sv57}, and much worse for 2D PTWs needed for virtualised systems. 
Our design may be more power efficient than a standard L2 TLB, since it can support up to 16$\times$ larger reach and thus may reduce costly PTWs.

\vspace{-0.4cm}
\section{Future Work}
\vspace{-0.3cm}
In this section we summarize further techniques with which SVNAPOT could improve system performance.

\vspace{-0.2cm}
\subsubsection{TLB Organization} We chose to evaluate VPN partitioning as a first approach to SVNAPOT due to the simplicity of the changes we had to make in the L2 TLB. Different approaches/organizations such as multiple TLB lookups or companion TLBs would yield different results\cite{ncppd} and may benefit architectures that support larger page sizes (128K, 256K etc.).
\vspace{-0.2cm}
\subsubsection{Contiguity} This is of critical importance, as without ample contiguity we cannot allocate large physical memory areas backed up by larger pages  \cite{capaging}. Increasing contiguity requires OS support. 

\subsubsection{Transparency} Larger allocations that can be backed up by contiguous regions could automatically be serviced by the kernel. Linux release v6.9 will support multi-size Transparent Huge Pages (mTHP) for ARM; RISC-V is not yet supported.

\vspace{-0.5cm}
\printbibliography 


\end{document}